\documentclass[twocolumn,showpacs,showkeys,preprintnumbers,superscriptaddress,amsmath,amssymb]{revtex4}

\usepackage[T1]{fontenc}
\usepackage[utf8]{inputenc}
\usepackage[bookmarksnumbered,bookmarksopen,colorlinks,citecolor=red,linkcolor=blue]{hyperref}
\usepackage{graphicx}
\usepackage{color}
\usepackage{latexsym,amssymb,amsmath,epsfig,bm,times,psfrag}
\usepackage{amsfonts,amsthm,amstext,amscd}
\usepackage{amsmath}
\usepackage{mathtools}
\usepackage{textcomp}
\usepackage{epstopdf}
\newcommand{\eq}[1]{Eq.~(\ref{#1})}

\begin{document}

\title{Impact of spin polarization on the QCD equation of state}

\author{De-Xian Wei}
\affiliation{School of Science, Guangxi University of Science and Technology, Liuzhou 545006, China}

\begin{abstract}
Spin polarization provides a novel probe of the rotational properties of the quark-gluon plasma formed in relativistic heavy-ion collisions. This work investigates the equation of state, particularly its transport and thermodynamic coefficients in noncentral O+O collisions, employing a parton distribution function that incorporates spin polarization induced by thermal vorticity. Within a kinetic theory framework, one finds that the magnitude of the squared speed of sound ($c_s^2$) is only weakly modified by spin polarization, whereas the specific shear viscosity ($\eta/s$), specific bulk viscosity ($\zeta/s$), and mean free path ($\lambda$) show substantial changes. When spin polarization is included, both $c_s^2$ and $\zeta/s$ develop a nonmonotonic dependence on the collision energy, with an inflection point near $\sqrt{s_{NN}}=27$ GeV, corresponding to an average parton chemical potential of $\langle\mu_p\rangle=0.021$ GeV. These results suggest that spin polarization may serve as a useful probe for constraining the effective equation of state of QCD matter.
\end{abstract}

\pacs{25.75.-q, 24.70.+s, 51.10.+y}
% PACS, the Physics and Astronomy Classification Scheme.

\keywords{Kinetic theory, spin polarization, equation of state, transport and thermodynamic coefficients}

\maketitle

\section{Introduction}
\label{introduction}
The study of nuclear matter governed by QCD is a central topic in high-energy physics. Relativistic heavy-ion collisions provide the primary experimental means to create and detect the quark-gluon plasma (QGP), which exhibits unique QCD properties. A subject of particular current interest is the observation of collective flow signals---typically associated with the QGP---in small systems such as proton-nucleus collisions, similar to those seen in larger nucleus-nucleus systems~\cite{CMS:2015yux,PHENIX:2018lia}. Whether such small systems genuinely produce QGP droplets, however, remains a matter of debate~\cite{Nagle:2018nvi}. Oxygen-oxygen (O-O) collisions, which may represent a minimal system capable of generating small QGP droplets, offer new opportunities to probe this question~\cite{Huss:2020dwe,Huss:2020whe,Eskola:2022vaf,Nijs:2025qxm,Singh:2025qab}. Recent studies have reported collective flow signals in O-O systems using A Multi-Phase Transport (AMPT) or UrQMD simulations~\cite{MenonKavumpadikkalRadhakrishnan:2025apq,Singh:2025pti}, although next-to-leading-order matrix-element analyses suggest that jet quenching---a hallmark of QGP formation---may be absent in such systems~\cite{Gebhard:2024flv}. To clarify the collective behavior and other features of O-O collisions, it is therefore essential to constrain their equation of state (EoS), particularly its transport and thermodynamic properties.

Over the past decades, substantial efforts have been devoted to determining the EoS of hot and dense QGP, particularly its transport and thermodynamic properties. First-principles calculations remain challenging owing to the fermion sign problem and the strongly coupled nature of the QGP~\cite{Troyer:2004ge,Altenkort:2022yhb}. A variety of effective models---including Yang-Mills theory~\cite{Haas:2013hpa}, Polyakov-Nambu-Jona-Lasinio (PNJL) models~\cite{Islam:2019tlo}, the color string percolation model~\cite{Sahu:2020mzo}, nonconformal holography~\cite{Grefa:2022sav}, the excluded-volume hadron resonance gas model~\cite{McLaughlin:2021dph}, the dynamical quasiparticle model ~\cite{Soloveva:2019xph}, effective kinetic theory~\cite{Gale:2020xlg}, and hydrodynamic models~\cite{Shen:2020mgh}---have been employed to study these properties. More recently, data-driven approaches such as machine learning and Bayesian inference have proven effective in extracting transport coefficients across system sizes~\cite{Nijs:2020roc,Bernhard:2019bmu,Moreland:2018gsh,JETSCAPE:2020mzn,JETSCAPE:2020shq,Heffernan:2023utr}, as summarized in Ref.~\cite{Achenbach:2023pba}. These analyses rely on many input parameters, underscoring the need for systematic studies to constrain real-time QGP properties. Present applications of machine learning and Bayesian inference are primarily guided by experimental data on conventional observables, with variations in initial conditions~\cite{Liu:2018xae}, temperature~\cite{Panday:2022sxb,Auvinen:2020mpc}, chemical potential~\cite{Sorensen:2021zme,He:2022kbc,He:2022yrk}, beam energy~\cite{Zabrodin:2020cfp}, and rotation~\cite{Sahoo:2023xnu,Rath:2024vyq}.

By contrast, relatively little attention has been paid to the role of spin polarization (SP), induced by vorticity or magnetic fields, as a constraint on QGP properties. Since SP naturally reflects the chiral imbalance of strongly interacting systems~\cite{Yang:2020ykt}, it serves as a promising probe of the EoS. Recent work has explored SP as a tool to investigate the QCD EoS~\cite{Singh:2021yba,Palermo:2024tza}, emphasizing its sensitivity to transport coefficients~\cite{Palermo:2024tza,Serenone:2021zef}. Moreover, the strong vorticity and magnetic fields generated in noncentral collisions can directly affect transport coefficients~\cite{Sahoo:2023xnu,Rath:2024vyq,Aung:2023pjf}. For example, it has been shown within second-order viscous hydrodynamics that the coupling of vorticity and viscosity significantly prolongs the lifetime of a rotating QGP~\cite{Sahoo:2023xnu}. In phenomenological analyses, when the spin relaxation time approaches the characteristic interaction timescale, spin degrees of freedom may alter the dynamical description of the medium. Thus, the mutual coupling between rotational effects and transport properties adds a further layer of complexity to our understanding of the space-time evolution of strongly interacting matter.

Nevertheless, these findings indicate that SP encodes information about the time-dependent transport and thermodynamic properties of the QGP. The primary objective of this work is to investigate, within noncentral O+O light-ion collisions, the influence of parton spin polarization on the effective transport and thermodynamic coefficients (TTCs). More broadly, exploring the EoS of strongly interacting matter through the kinetic theory of SP provides a promising theoretical framework.

\section{Spin polarization in kinetic theory}
\label{method}

The spin-dependent phase-space distribution function in a near-equilibrium state can be parametrized as~\citep{Weickgenannt:2020aaf,Florkowski:2019gio,Becattini:2013fla}
\begin{eqnarray}\label{disf:101}
& & \tilde{f}_{mn}[r(\tau),p(\tau),s(\tau)]   \nonumber\\
&=& f[r(\tau),p(\tau)] \times \left[\delta_{mn}+\Pi[r(\tau),p(\tau)]\sigma_{mn}\right]\,.
\end{eqnarray}
Here, $m,n=1,2$ are spin indices, $f[r(\tau),p(\tau)]$ is the spinless distribution function, $s(\tau)$ denotes the spin degree of freedom in phase space, $\sigma_{mn}$ is a three-vector composed of Pauli matrices, and $\Pi[r(\tau),p(\tau)]$ represents the spin-polarization three-vector.

Assuming small vorticity within local thermodynamic equilibrium, a linear relation emerges between the averaged spin polarization vector and the thermal vorticity. For spin-$1/2$ particles, the polarization vector satisfies $\Pi^{\mu}\approx \varpi^{\mu}[2s(s+1)/3]$. A similar relation between $\Pi^{\mu}$ and $\varpi^{\mu}$ is discussed in Refs.~\citep{Becattini:2013fla,Becattini:2016gvu}.

For SP studies, it is convenient to express thermal vorticity in tensor form: $\varpi_{\rho\sigma}=1/2\,(\partial_{\sigma}\beta_{\rho}-\partial_{\rho}\beta_{\sigma})$, while the thermal vorticity vector is given by $\varpi^{\mu}=1/2\,\epsilon^{\mu\rho\sigma\tau}\,u_{\rho}\,\partial_{\sigma}\beta_{\tau} =-1/2\,\epsilon^{\mu\rho\sigma\tau}\,\varpi_{\rho\sigma}\,u_{\tau},$ where $\beta_{\rho}=u_{\rho}/T$, with $u_{\rho}$ the four-velocity of the fluid and $T$ the temperature.

To highlight the role of SP, I employ kinetic theory to calculate the energy-momentum tensor. Including the spin measure, the SP-dependent energy-momentum tensor is defined as
\begin{eqnarray}\label{tens:101}
\tilde{T}^{\mu\nu}(\tau,r) &=& \frac{1}{2}\int\frac{d^{3}p}{(2\pi)^{3}}\frac{p^{\mu}p^{\nu}}{p^{0}}\int dS\tilde{f}[r(\tau),p(\tau),s(\tau)] \nonumber\\
&=& \int\frac{d^{3}p}{(2\pi)^{3}}\frac{p^{\mu}p^{\nu}}{p^{0}}f[r(\tau),p(\tau)]  \nonumber\\
&\times& \frac{1}{2}\sum_{m,n=1}^{2}\int \left\{\delta_{mn}+\Pi[r(\tau),p(\tau)]\sigma_{mn}\right\}dS   \nonumber\\
&=& \int\frac{d^{3}p}{(2\pi)^{3}}\frac{p^{\mu}p^{\nu}}{p^{0}}f[r(\tau),p(\tau)]  \nonumber\\
&\times& \frac{1}{2}\sum_{m=1}^{2}\int \left\{\delta_{mm}+\Pi[r(\tau),p(\tau)]\sigma_{mm}\right\}dS \,.
\end{eqnarray}

The invariant spin measure can be parameterized as~\citep{Florkowski:2018fap}
\begin{eqnarray}\label{tens:102}
& & \frac{1}{2}\sum_{m=1}^{2}\int \left\{\delta_{mm}+\Pi[r(\tau),p(\tau)]\sigma_{mm}\right\}dS   \nonumber\\
&\approx& 2-\frac{\, \sinh{\chi\,\Pi[r(\tau),p(\tau)]}}{\chi\,\Pi[r(\tau),p(\tau)]}\,,
\end{eqnarray}
with
\begin{eqnarray}\label{tens:103}
& & \int \langle\cdot\cdot\cdot\rangle dS=\frac{M}{\pi\chi}\int \langle\cdot\cdot\cdot\rangle d^{4}s\delta(s\cdot s+\chi^{2})\delta(p\cdot s) \,,
\end{eqnarray}
where $\chi^{2}={3}/{4}$,  $M$ is the particle's mass, and the factor $1/2$ denotes the normalization of the sum of particles and antiparticles (this work does not distinguish between particles and antiparticles, as the chemical potential is not introduced in this framework). The distribution function sums over spin-up and spin-down particles. When spin is treated as an independent variable, disregarding contributions from thermal vorticity, the result of integrating the spin measure is $\int dS= 2$~\cite{Florkowski:2018fap}.

For comparison, the spinless energy-momentum tensor is

\begin{eqnarray}\label{tens:104}
T^{\mu\nu}(\tau,r) &=& \int\frac{d^{3}p}{(2\pi)^{3}}\frac{p^{\mu}p^{\nu}}{p^{0}}f[r(\tau),p(\tau)].
\end{eqnarray}

Here $d^{3}p/(2\pi)^{3}/p^{0}$ is the Lorentz-invariant integration measure in momentum space, with $p^{0}=\sqrt{m^{2}+p^{2}}$ the on-shell particle energy.

By evaluating the energy–momentum tensor, one can extract the TTCs of interest in QGP physics, such as the specific shear viscosity $\eta/s$, the specific bulk viscosity $\zeta/s$, and the mean free path $\lambda$. The first-order, spinless TTCs for massive particles in kinetic theory are defined as

\begin{eqnarray}\label{tran:101}
\eta/s &=& \frac{1}{15(\varepsilon+P)}\int\frac{d^{3}p}{(2\pi)^{3}}\frac{p^{4}}{p^{0}}f(1-f) \nonumber\\
\zeta/s &=& 15\eta/s\left(\frac{1}{3}-c_{s}^{2}\right)^{2}  \nonumber\\
&=& \frac{1}{(\varepsilon+P)}\left(\frac{1}{3}-c_{s}^{2}\right)^{2} \int\frac{d^{3}p}{(2\pi)^{3}}\frac{p^{4}}{p^{0}}f(1-f)  \nonumber\\
\lambda &\approx& \frac{5}{T}\eta/s     \nonumber\\
&=& \frac{1}{3(\varepsilon+P)T}\int\frac{d^{3}p}{(2\pi)^{3}}\frac{p^{4}}{p^{0}}f(1-f)\,.
\end{eqnarray}

Here, the local temperature is defined as $T=(\pi^{2}\varepsilon/72)^{1/4}$~\cite{Zhang:2015tta}. The energy density and pressure follow from \eq{tens:104}: $\varepsilon=T^{00},~P=\sqrt{(T^{11})^{2}+(T^{22})^{2}+(T^{33})^{2}}$. The squared speed of sound is given by the derivative of the pressure with respect to the energy density at constant entropy: $c_{s}^{2}=(\partial P/\partial \varepsilon)_{s}$.

Similarly, the first-order SP-dependent TTCs for massive particles are defined as
\begin{eqnarray}\label{tran:102}
\tilde{\eta}/\tilde{s} &=& \frac{1}{15(\tilde{\varepsilon}+\tilde{P})}\int\frac{d^{3}p}{(2\pi)^{3}}\frac{p^{4}}{p^{0}}dS\,\tilde{f}(1-\tilde{f}) \nonumber\\
\tilde{\zeta}/\tilde{s} &=& \frac{1}{(\tilde{\varepsilon}+\tilde{P})}\left(\frac{1}{3}-\tilde{c}_{s}^{2}\right)^{2} \int\frac{d^{3}p}{(2\pi)^{3}}\frac{p^{4}}{p^{0}}dS\,\tilde{f}(1-\tilde{f})  \nonumber\\
\tilde{\lambda} &\approx& \frac{1}{3(\tilde{\varepsilon}+\tilde{P})T}\int\frac{d^{3}p}{(2\pi)^{3}}\frac{p^{4}}{p^{0}}dS\,\tilde{f}(1-\tilde{f})\,,
\end{eqnarray}
where $\tilde{\varepsilon}=\tilde{T}^{00},~\tilde{P}=\sqrt{(\tilde{T}^{11})^{2}+(\tilde{T}^{22})^{2}+(\tilde{T}^{33})^{2}}$, and $\tilde{c}_{s}^{2}=(\partial \tilde{P}/\partial \tilde{\varepsilon})_{s}$. Both the vorticity tensor and the viscosity coefficients are dissipative quantities governed by the relaxation time.

In this work, the TTCs are investigated in O+O collisions with an impact parameter $b=3$ fm, using the AMPT model~\citep{Lin:2004en}. No multiplicity filter is applied; instead, the impact parameter is fixed, yielding a broad range of multiplicities from low to high.

The default AMPT collision-probability parameters are used: $a=0.5$, $b=0.9$ GeV$^{-2}$, $\alpha_{s}=0.33$, and $\mu=3.2$ fm$^{-1}$. The AMPT model includes transport processes that simulate parton interactions and incorporates a natural viscosity determined by the scattering cross section, as formulated in the Israel-Stewart or Chapman-Enskog framework~\citep{MacKay:2022uxo,Zhang:2023lzv}.

To investigate the effect of SP on transport and thermodynamic properties, I reextract the TTCs within the kinetic theory framework, as defined in Eqs.~(\ref{tran:101}) and (\ref{tran:102}).
Spin polarization is included only in the partonic stage and is assumed to arise solely from thermal vorticity, which reflects the system’s orbital angular momentum, while possible contributions from magnetic fields, thermal shear, and viscous effects are neglected.
This work adopts natural units with $k_{B}=c=\hbar=1$.

\section{Numerical results}
\label{results}
Having introduced the kinetic theory framework, I now present numerical results for the TTCs. A more detailed numerical analysis will be presented elsewhere.
\begin{figure}[tp]
\includegraphics[width=0.480\textwidth]{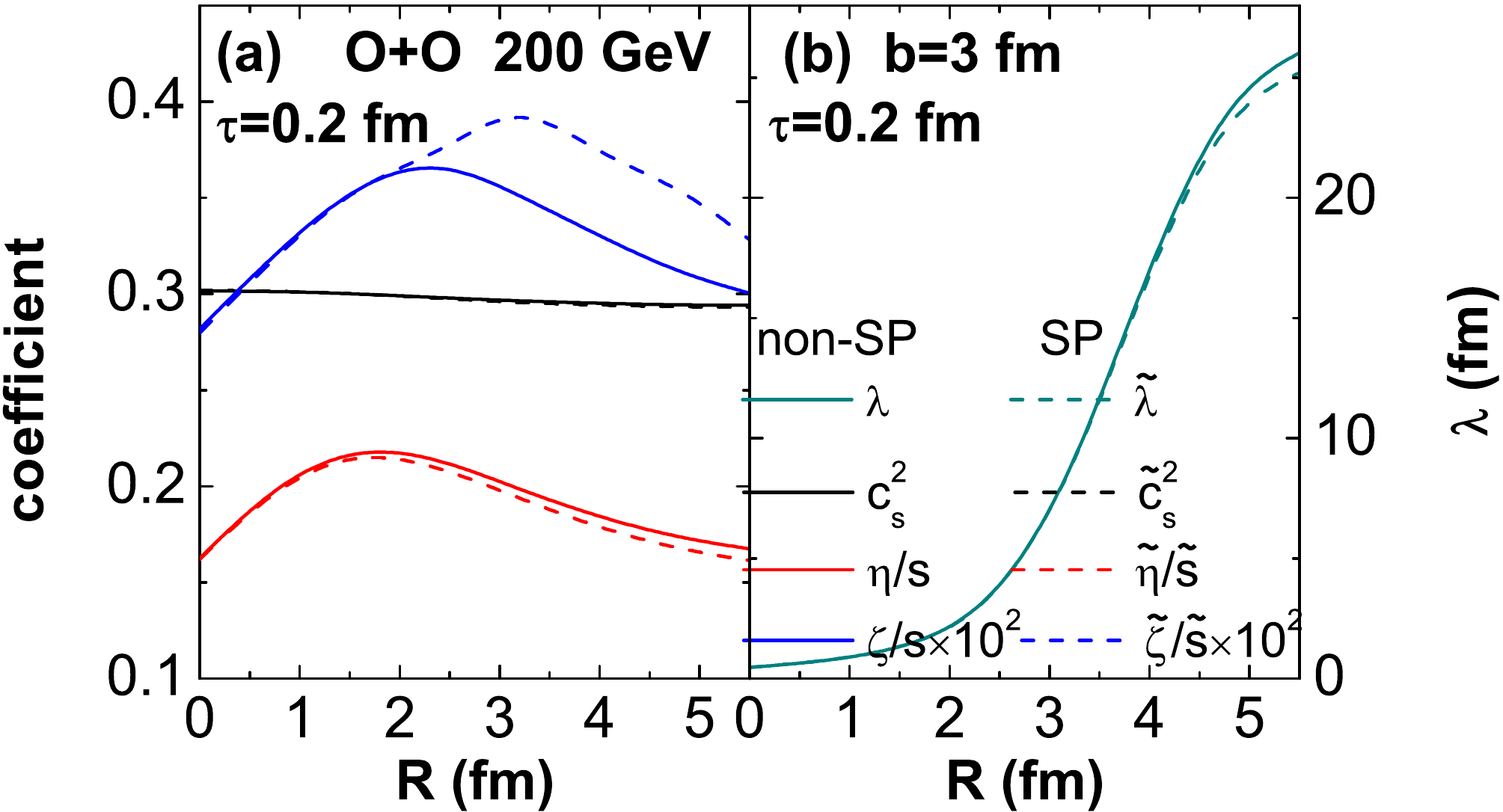}
\caption{(Color online) Comparison of SP and non-SP results: (a) squared speed of sound $c_{s}^{2}$, specific shear viscosity $\eta/s$, and specific bulk viscosity $\zeta/s$ and (b) mean free path $\lambda$, shown as functions of radius for O+O collisions at $\sqrt{s_{NN}}=200$ GeV. Results correspond to proper time $\tau=0.2$ fm.}
\label{fig1}
\end{figure}

\begin{figure}[tp]
\includegraphics[width=0.480\textwidth]{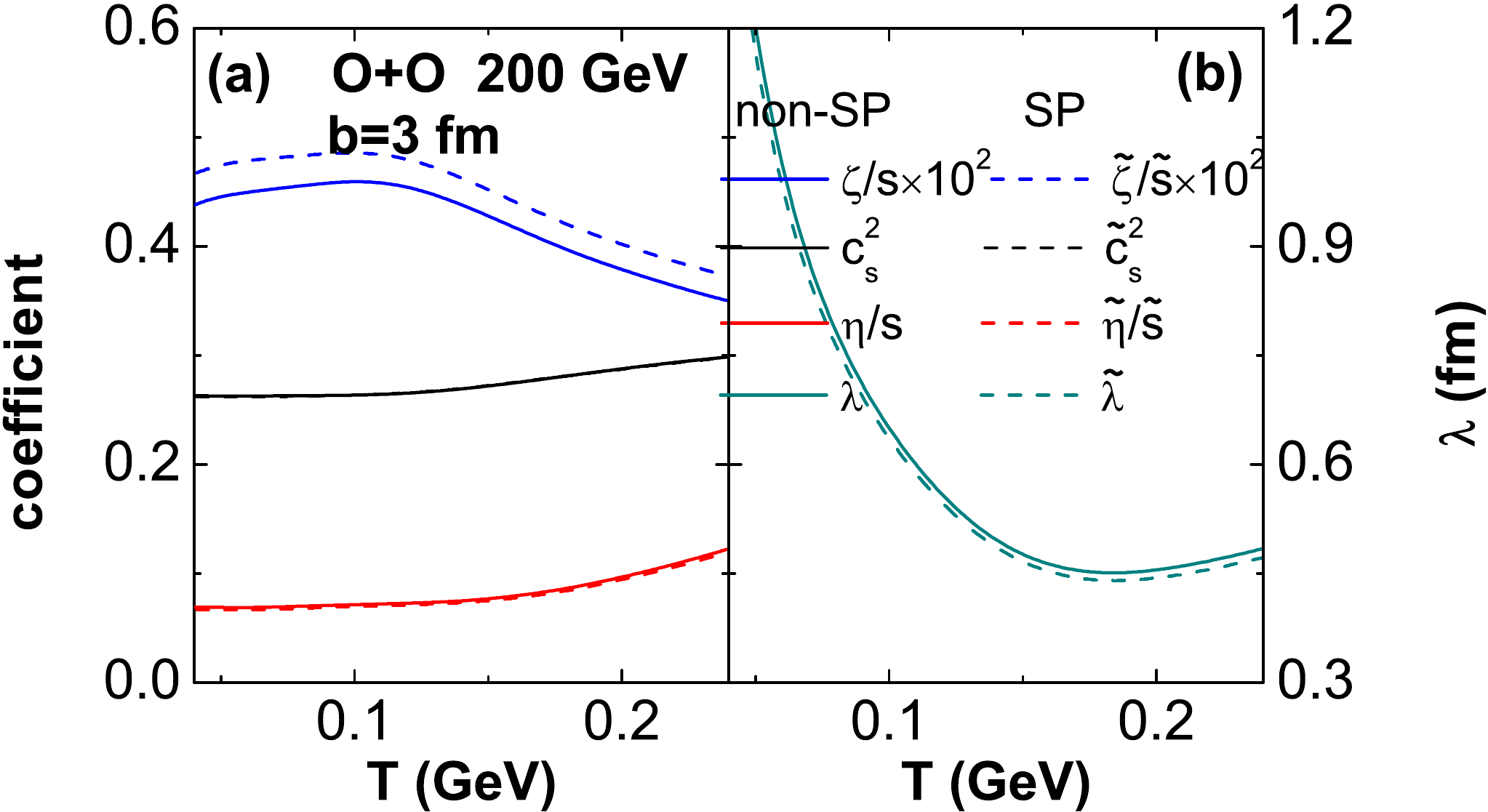}
\caption{(Color online) Comparison of SP and non-SP results: (a) squared speed of sound $c_{s}^{2}$, specific shear viscosity $\eta/s$, and specific bulk viscosity $\zeta/s$ and (b) mean free path $\lambda$, shown as functions of temperature for O+O collisions at $\sqrt{s_{NN}}=200$ GeV. Results are integrated over proper time from $\tau=0.2$ to $5$ fm.}
\label{fig2}
\end{figure}

Figure~\ref{fig1} shows the squared speed of sound $c_{s}^{2}$, specific shear viscosity $\eta/s$, specific bulk viscosity $\zeta/s$, and mean free path $\lambda$ as functions of radius for O+O collisions at $\sqrt{s_{NN}}=200$ GeV, evaluated at proper time $\tau=0.2$ fm. All coefficients except $c_{s}^{2}$ exhibit significant sensitivity to SP.

Vorticity defines a noninertial reference frame in which physical quantities acquire noninertial corrections. Thus, SP, as a nondissipative effect, modifies the transport properties of the system in such a frame~\cite{Becattini:2013fla}. In Fig.~\ref{fig1}, the splitting between SP and non-SP results increases with radius, reflecting the quadrupole structure of thermal vorticity in coordinate space~\cite{Wei:2018zfb}. Hence, the influence of SP on the TTCs is volume dependent.

SP affects shear and bulk viscosities in opposite ways.
Across the radius range, SP suppresses $\eta/s$. In contrast, its effect on $\zeta/s$ changes sign: $\zeta/s$ is suppressed at low radius but enhanced at high radius, with a crossover near $R \approx 2$ fm.
Shear viscosity, which characterizes momentum transport, is mainly influenced by Coriolis forces in a rotating system~\cite{Aung:2023pjf}. Bulk viscosity, in contrast, describes pressure variations due to expansion or compression of the fluid. As a result, SP contributions to these two coefficients show opposite trends. The mean free path $\lambda$, obtained directly from $\eta/s$, qualitatively follows its behavior.
Note that the results at $\sqrt{s_{NN}}=200$ GeV correspond to the early stage of system evolution, when the coefficients are most strongly modified. At lower collision energies, $\eta/s$ and $\lambda$ show only a weak time dependence, as illustrated in Fig.~\ref{fig:coefftime}. Time-dependent splitting effects are discussed in the appendix.

Figure~\ref{fig2} presents $c_{s}^{2}$, $\eta/s$, $\zeta/s$, and $\lambda$ as functions of temperature for O+O collisions at $\sqrt{s_{NN}}=200$ GeV, integrated over the proper time interval $\tau=0.2$--$5$ fm. As in Fig.~\ref{fig1}, all coefficients except $c_{s}^{2}$ are strongly affected by SP.
Across the temperature range, SP suppresses $\eta/s$, while it enhances $\zeta/s$.
As with $\eta/s$, the SP contribution to $\lambda$ remains suppressive, consistent with its direct relation to $\eta/s$.
These results are consistent with the conclusion from Fig.~\ref{fig1}.

\begin{figure}[tp]
\includegraphics[width=0.480\textwidth]{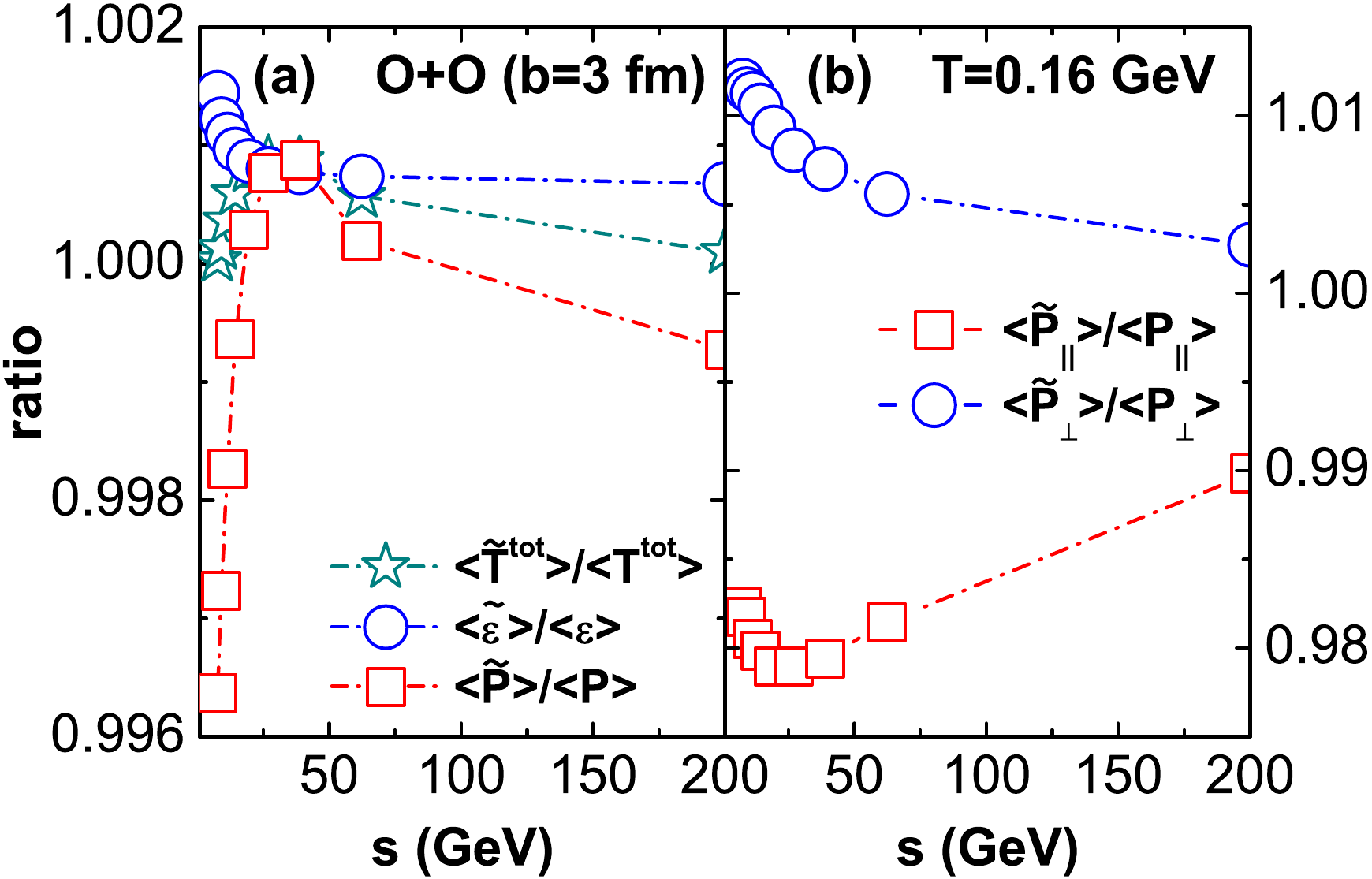}
\caption{(Color online) Comparison of SP and non-SP results for event-averaged quantities: (a) ratios $\langle \tilde{T}^{\mathrm{tot}}\rangle/\langle T^{\mathrm{tot}}\rangle$, $\langle \tilde{\varepsilon}\rangle/\langle \varepsilon\rangle$, and  $\langle \tilde{P}\rangle/\langle P\rangle$ and (b) ratios $\langle \tilde{P}_{\parallel}\rangle/\langle P_{\parallel}\rangle$ and $\langle \tilde{P}_{\perp}\rangle/\langle P_{\perp}\rangle$, shown as functions of energy for O+O collisions at $T=0.16$ GeV. Results are integrated over proper time from $\tau=0.2$ to 5 fm.}
\label{fig3}
\end{figure}

\begin{figure}[tp]
\includegraphics[width=0.480\textwidth]{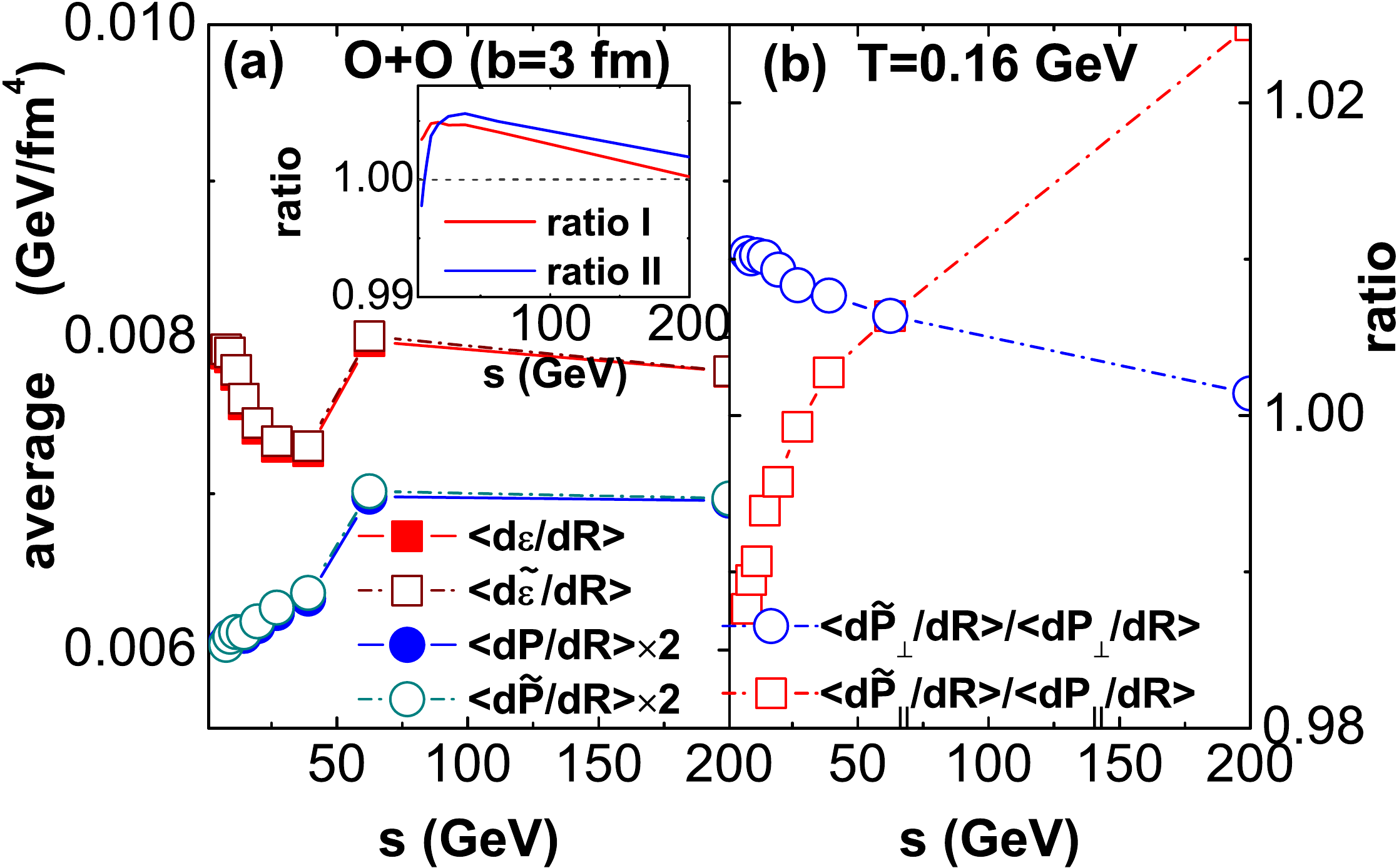}
\caption{(Color online) Comparison of SP and non-SP results for event-averaged quantities:
(a) gradient of energy density $\langle d\varepsilon/dR\rangle$ and gradient of pressure $\langle dP/dR\rangle$ and
(b) ratios $\langle d\tilde{P}_{\parallel}/dR\rangle/\langle dP_{\parallel}/dR\rangle$ and $\langle d\tilde{P}_{\perp}/dR\rangle/\langle dP_{\perp}/dR\rangle$, shown as functions of energy for O+O collisions at $T=0.16$ GeV.
Results are integrated over proper time from $\tau=0.2$ to 5 fm. The inset shows two ratios: (I) $\langle d\tilde{\varepsilon}/dR\rangle/\langle d\varepsilon/dR\rangle$ and (II) $\langle d\tilde{P}/dR\rangle/\langle dP/dR\rangle$.}
\label{fig4}
\end{figure}

\begin{figure}[tp]
\includegraphics[width=0.480\textwidth]{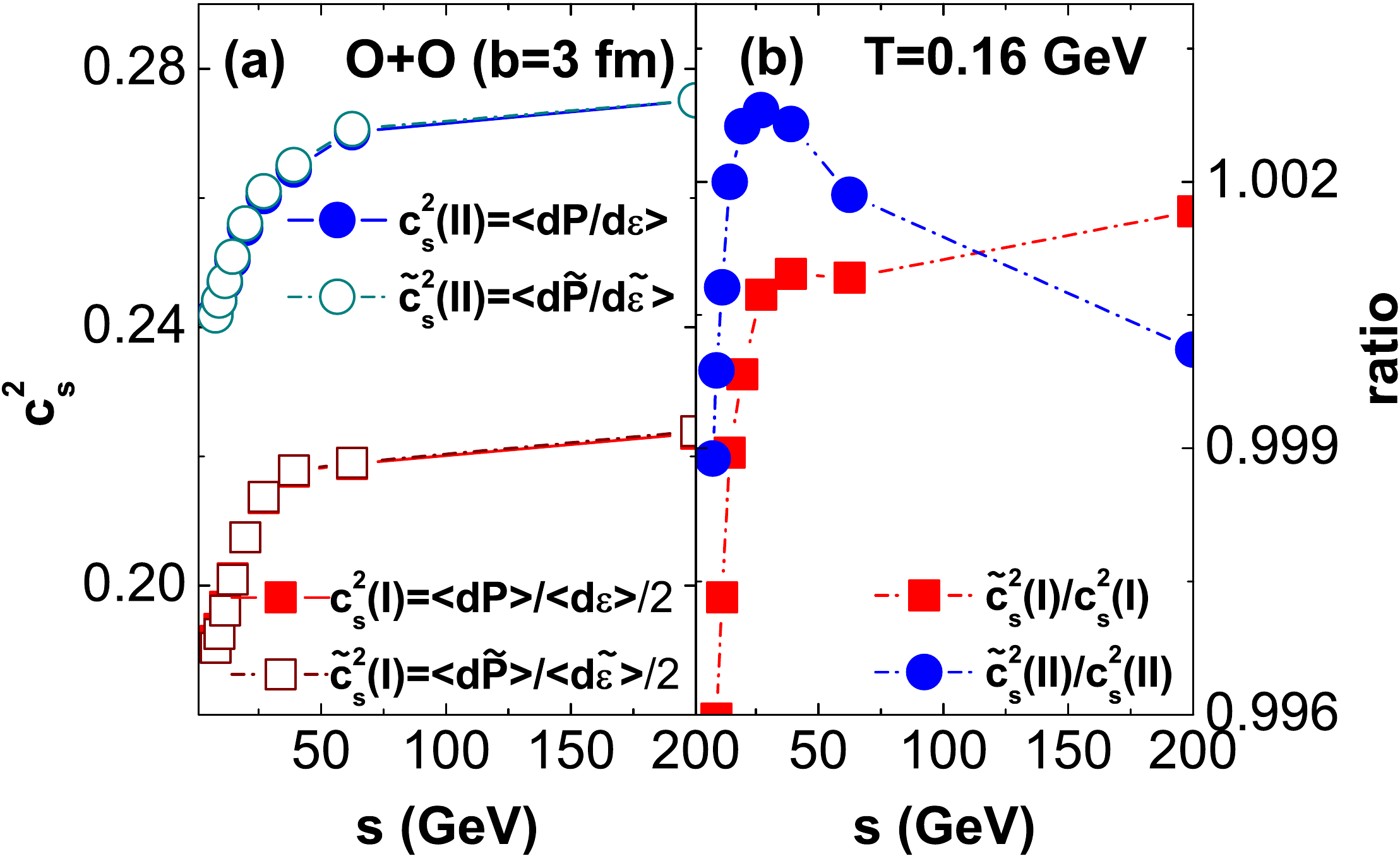}
\caption{(Color online) Comparison of two event-averaging methods: case I and case II.
(a) Squared speed of sound, comparing SP and non-SP results, shown as a function of energy for O+O collisions at $T=0.16$ GeV.
(b) Ratio $\tilde{c}_{s}^{2}/c_{s}^{2}$, also shown as a function of energy for O+O collisions at $T=0.16$ GeV.
Results are integrated over proper time from $\tau=0.2$ to 5 fm.}
\label{fig5}
\end{figure}

To clarify the nonmonotonic behavior of SP in Fig.~\ref{fig1} with respect to the TTCs, I performed an event-averaged analysis of the energy-momentum tensor and its spatial gradients. The results are shown in Figs.~\ref{fig3}--\ref{fig5}. All quantities are integrated over the proper time from $\tau=0.2$ to 5 fm.

Figure~\ref{fig3} presents ratios of event-averaged quantities:
(a) $\langle \tilde{T}^{\mathrm{tot}}\rangle/\langle T^{\mathrm{tot}}\rangle$, $\langle \tilde{\varepsilon}\rangle/\langle \varepsilon\rangle$, and  $\langle \tilde{P}\rangle/\langle P\rangle$ and (b) $\langle \tilde{P}_{\parallel}\rangle/\langle P_{\parallel}\rangle$ and $\langle \tilde{P}_{\perp}\rangle/\langle P_{\perp}\rangle$, shown as functions of collision energy at $T=0.16$ GeV.
Here the energy-momentum tensor is
$\langle T^{\mathrm{tot}}\rangle = \langle\sqrt{(T^{00})^{2}+(T^{11})^{2}+(T^{22})^{2}+(T^{33})^{2}}\rangle$,
energy density is $\langle\varepsilon\rangle = \langle T^{00}\rangle$,
pressure is $\langle P\rangle = \langle\sqrt{(T^{11})^{2}+(T^{22})^{2}+(T^{33})^{2}}\rangle$,
transverse pressure is  $\langle P_{\parallel}\rangle = \langle\sqrt{(T^{11})^{2}+(T^{22})^{2}}\rangle$,
and longitudinal pressure is $\langle P_{\perp}\rangle = \langle T^{33}\rangle$.

Figure~\ref{fig4} shows the corresponding gradients:
$\langle d\varepsilon/dR\rangle$, $\langle d\tilde{\varepsilon}/dR\rangle$,
$\langle dP/dR\rangle$,
and $\langle d\tilde{P}/dR\rangle$,
and ratios $\langle d\tilde{P}_{\parallel}/dR\rangle/\langle dP_{\parallel}/dR\rangle$ and $\langle d\tilde{P}_{\perp}/dR\rangle/\langle dP_{\perp}/dR\rangle$,
as functions of collision energy at $T=0.16$ GeV.

As indicated in Figs.~\ref{fig3} and \ref{fig4}, the SP--non-SP splitting is significantly stronger for the quantities. These splitting effects are generally monotonic in energy,
except for $\langle T^{\mathrm{tot}}\rangle$, $\langle P\rangle$, and $\langle P_{\parallel}\rangle$.
The SP suppresses the $\langle T^{\mathrm{tot}}\rangle$ and $\langle P\rangle$ at lower and higher energy but enhances them at middle energy.
The SP suppresses the $\langle P_{\parallel}\rangle$ in the entire energy range, with a valley near 19.6 GeV.

The microscopic origin of this behavior lies in nonlocal multiple scattering~\cite{Zhang:2019xya,Sheng:2021kfc}. The energy levels of fermions shift with the magnitude of the vorticity, altering the density of states. Consequently, the contribution of SP to $\langle P\rangle$ arises from the redistribution of the energy density induced by multiple scattering. As shown in Fig.~\ref{fig:gradenergy}, SP affects $\langle P\rangle$  differently depending on the collision energy:
In high-energy systems, the suppression of SP on $\langle P\rangle$ dominates at early times, whereas the suppression of SP becomes more pronounced over time in low-energy systems. This implies that low-energy systems require a longer thermalization period. In other words, SP prolongs the medium lifetime, consistent with Ref.~\cite{Sahoo:2023xnu}.

In Fig.~\ref{fig5}, two different event averaging methods are compared.
In case I, the event average is computed first, and the ratio is taken afterward.
In case II, the ratio is evaluated for each event and then averaged.
As shown in the figure, both methods yield the same qualitative behavior of SP; they differ only in the magnitude of the sound velocity.

Figure~\ref{fig5}(a) shows that, in both cases, the squared sound velocity increases monotonically with the collision energy.
In contrast, Fig.~\ref{fig5}(b) demonstrates that the SP contribution to $c_{s}^{2}$ exhibits differential energy dependence:
In case I, the effect of SP on $c_{s}^{2}$ is monotonic; in case II, the effect of SP on $c_{s}^{2}$ is nonmonotonic, with a peak near 27 GeV. The results are supported by Fig.~\ref{fig4}(a).
Note that the calculation of the TTCs presented here follows the event-averaging method of case II in Fig.~\ref{fig5}.

\begin{figure*}[t]
\begin{center}
\includegraphics[width=0.450\textwidth]{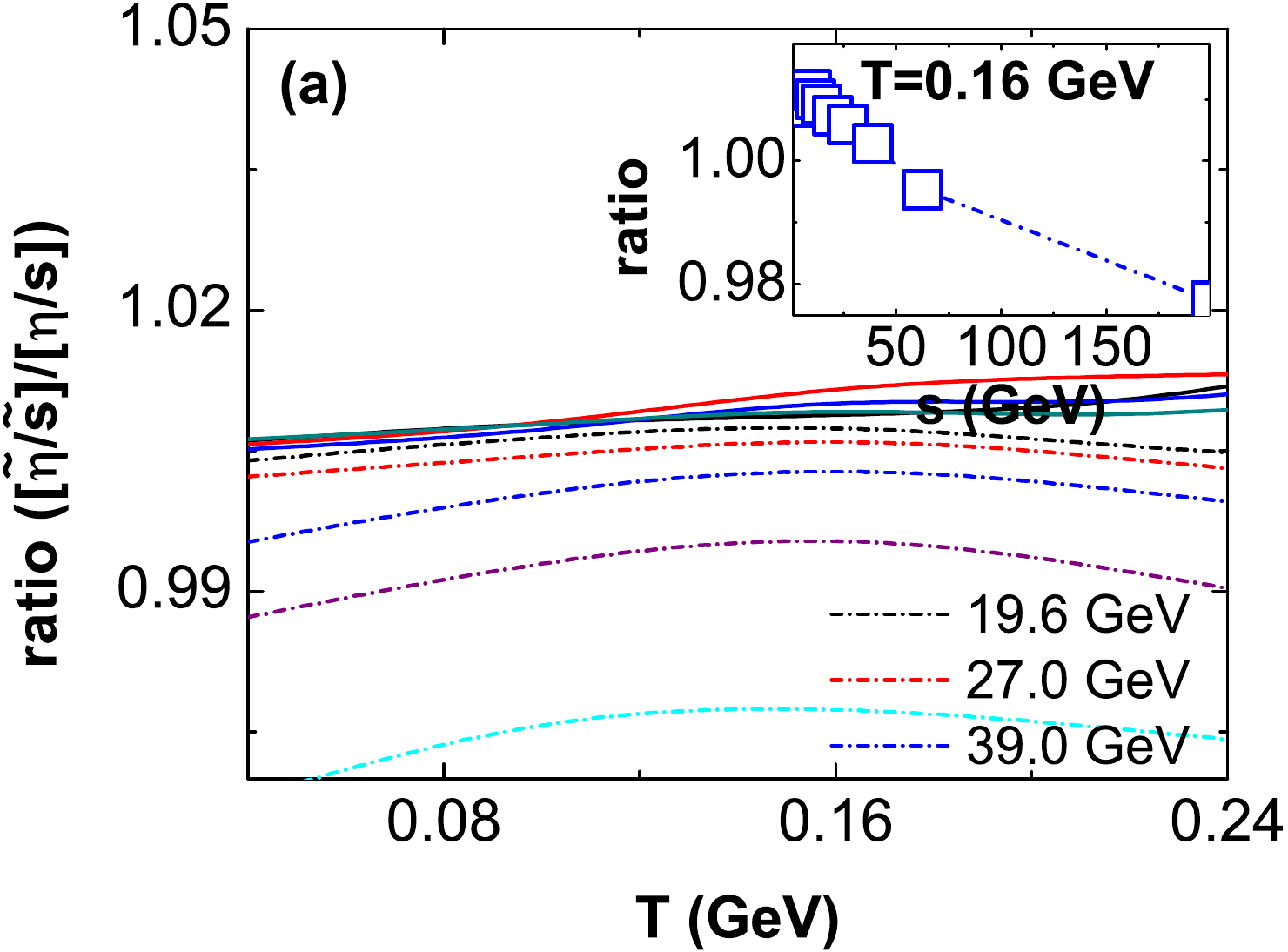}
\hspace{0.20cm}
\includegraphics[width=0.450\textwidth]{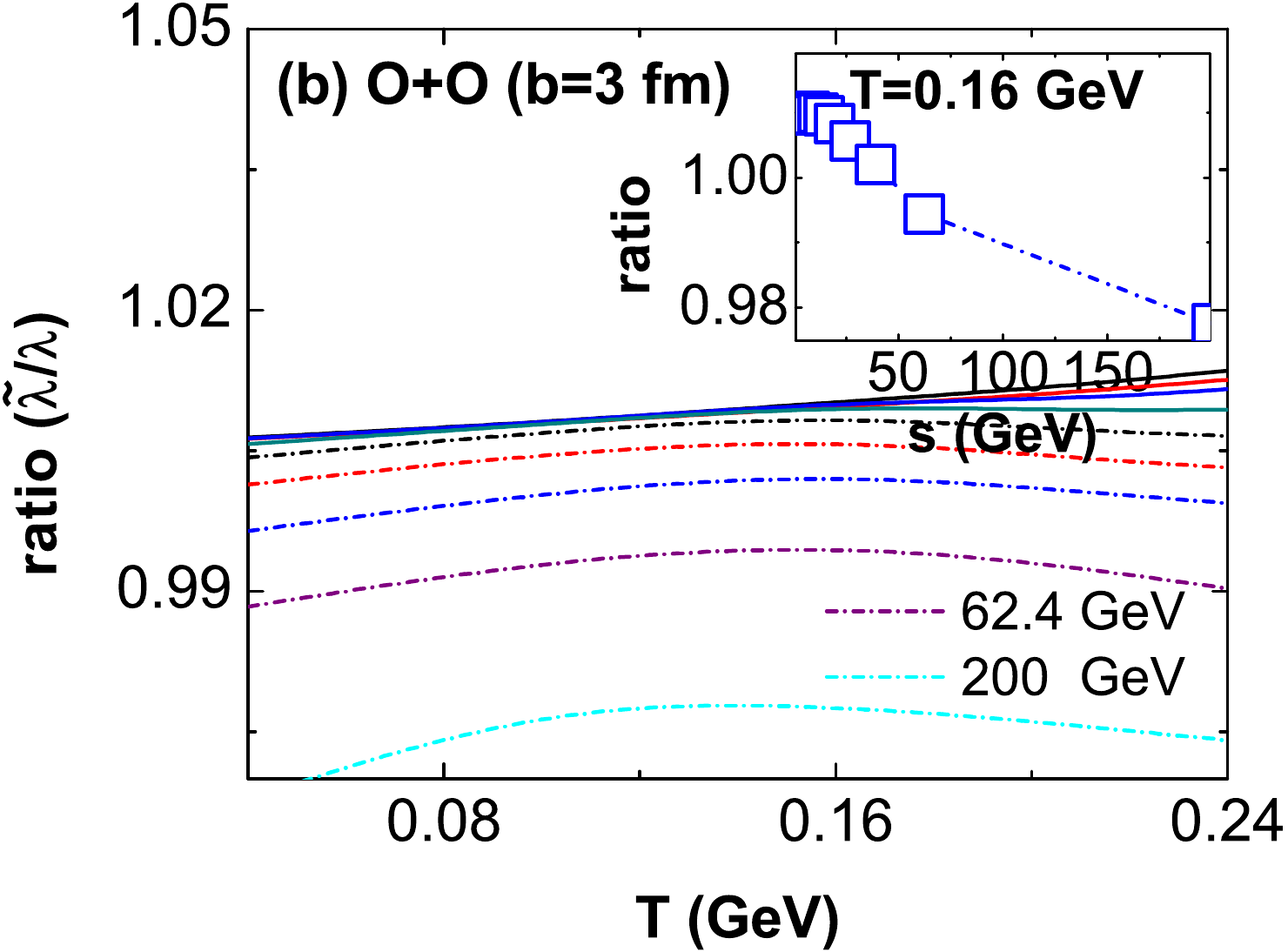}\\
\includegraphics[width=0.450\textwidth]{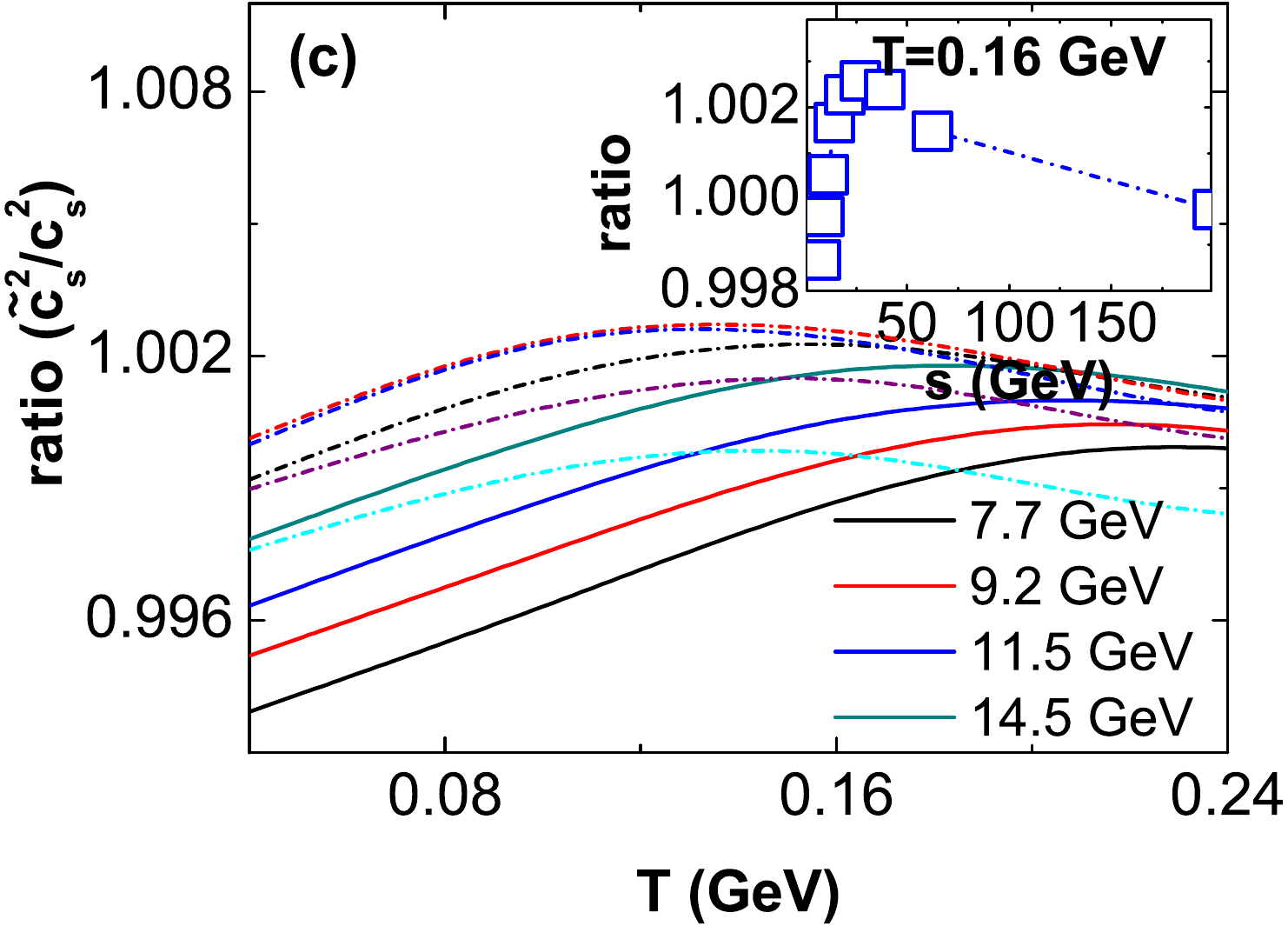}
\vspace{0.20cm}
\includegraphics[width=0.450\textwidth]{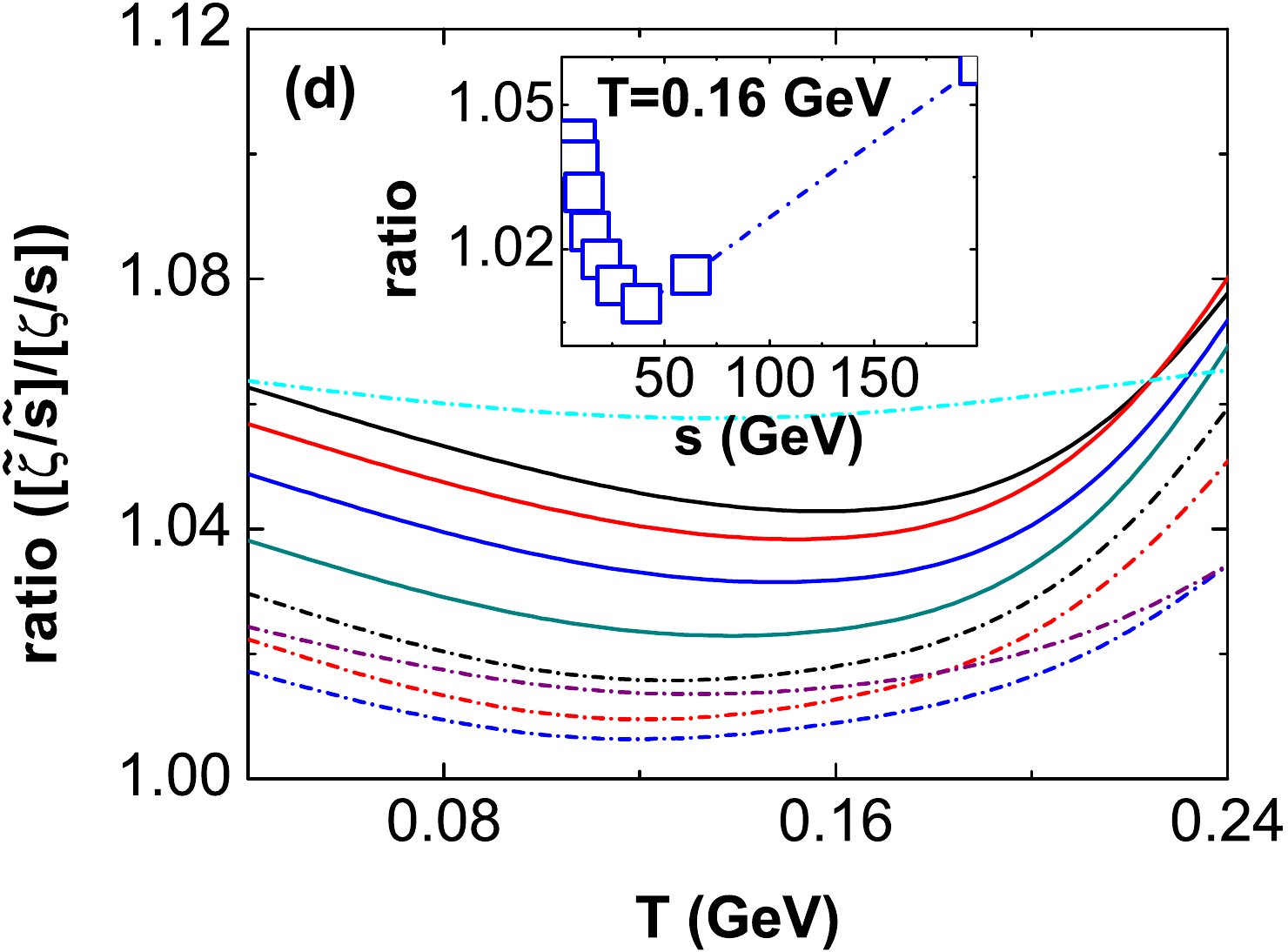}
\caption{(Color online)
Top panels: The ratios $(\tilde{\eta}/\tilde{s})/(\eta/s)$ and $\tilde{\lambda}/\lambda$ as a function of temperature, integrated over proper time from $\tau=0.2$ to 5 fm.
Bottom panels: The same distribution for $\tilde{c}_{s}^{2}/c_{s}^{2}$ and $(\tilde{\zeta}/\tilde{s})/(\zeta/s)$.
Results correspond to O+O collisions at $\sqrt{s_{NN}}=7.7$--200 GeV.}
\label{fig6}
\end{center}
\end{figure*}

To further test the conclusions from Fig.~\ref{fig5}, Fig.~\ref{fig6} shows the $T\!-\!\sqrt{s_{NN}}$ distributions of the ratios $\tilde{c}_{s}^{2}/c_{s}^{2}$, $(\tilde{\eta}/\tilde{s})/(\eta/s)$, $(\tilde{\zeta}/\tilde{s})/(\zeta/s)$, and $\tilde{\lambda}/\lambda$.

The top panels display the $T\!-\!\sqrt{s_{NN}}$ distributions of $\eta/s$ and $\lambda$, comparing SP and non-SP results.
The bottom panels present the $T\!-\!\sqrt{s_{NN}}$ distributions of the same ratios, for $\tilde{c}_{s}^{2}/c_{s}^{2}$ and $(\tilde{\zeta}/\tilde{s})/(\zeta/s)$.
At $T =0.16$ GeV, both $(\tilde{\eta}/\tilde{s})/(\eta/s)$ and $\tilde{\lambda}/\lambda$ decrease monotonically with increasing collision energy,
while both $\tilde{c}_{s}^{2}/c_{s}^{2}$ and $(\tilde{\zeta}/\tilde{s})/(\zeta/s)$ show nonmonotonic behavior, with a peak around 27 GeV.
The nonmonotonicity of $(\tilde{\zeta}/\tilde{s})/(\zeta/s)$ is linked to the squared speed of sound $c_{s}^{2}$, which drives the reversed trend observed in $\tilde{c}_{s}^{2}/c_{s}^{2}$.
Note that the behavior follows from the definition of the squared sound velocity, $c_{s}^{2}=(\partial P/\partial \varepsilon)_{s}$, as illustrated in Fig.~\ref{fig5}.
In Fig.~\ref{fig6}, all TTCs display clear temperature scaling for $T \lesssim 0.16$ GeV, where they depend only on collision energy and become independent of temperature.
SP characterizes a nondissipative quantity in local thermal equilibrium, while specific viscosities are dissipative and relevant to nonequilibrium evolution.
Thus, the contribution of SP to the viscosity coefficients may reflect the temperature scale invariance for $T \lesssim 0.16$ GeV.
This invariance implies that the system reaches thermal equilibrium in this regime, where the contribution of the SP to the dissipation coefficients remains constant.
The results are consistent with, and further support, the conclusions from Fig.~\ref{fig5}.

\begin{figure}[tp]
\begin{center}
\includegraphics[width=0.480\textwidth]{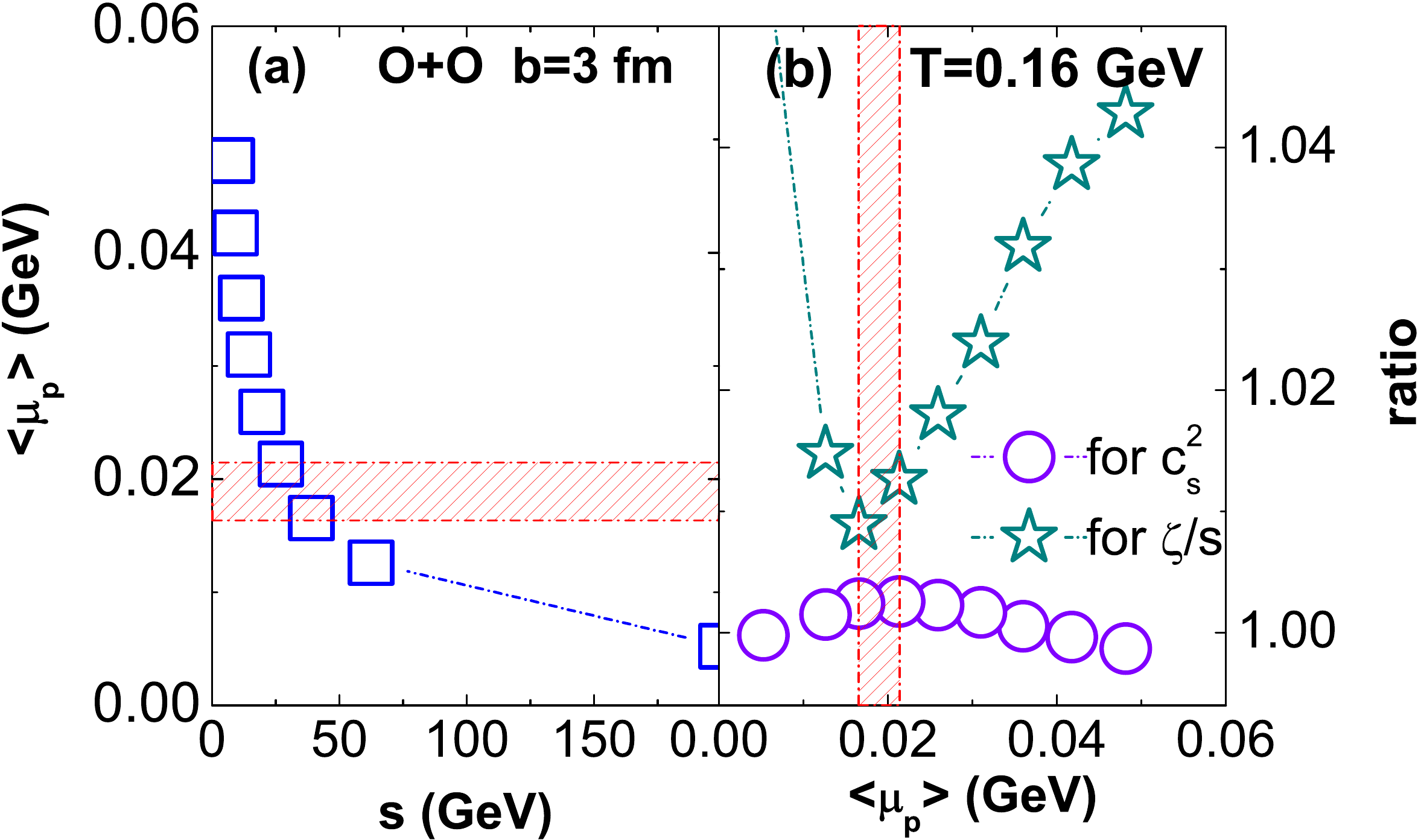}
\caption{(Color online)
Left panel: Average parton chemical potential $\langle\mu_{p}\rangle$ as a function of collision energy.
Right panel: Ratios $\tilde{c}_{s}^{2}/c_{s}^{2}$ and $(\tilde{\zeta}/\tilde{s})/(\zeta/s)$ as functions of the average parton chemical potential $\langle\mu_{p}\rangle$.}
\label{fig7}
\end{center}
\end{figure}

To investigate QCD properties at finite chemical potential, I introduce the averaged parton chemical potential, defined as
\begin{eqnarray}\label{chem:101}
\langle\mu_{p}\rangle &=& \frac{\int \mu_{p}(\tau)\, d\tau}{\int d\tau},
\end{eqnarray}
where $\mu_{p}(\tau)=(T_{c}/2)\,\log[N_{p}(\tau)/N_{\bar{p}}(\tau)]$ is computed for a single collision event, following Ref.~\cite{Yu:2014epa}.
The particle number is defined as $N_{p}(\tau)=N_{u}(\tau)+N_{d}(\tau)+N_{s}(\tau)$ and the antiparticle number as $N_{\bar{p}}(\tau)=N_{\bar{u}}(\tau)+N_{\bar{d}}(\tau)+N_{\bar{s}}(\tau)$, with subscripts denoting quark flavors.
The critical temperature is fixed at $T_{c}=0.16$ GeV.

Figure~\ref{fig7}(a) shows $\langle\mu_{p}\rangle$, which decreases monotonically with increasing collision energy.
Figure~\ref{fig7}(b) presents the ratios $\tilde{c}_{s}^{2}/c_{s}^{2}$ and $(\tilde{\zeta}/\tilde{s})/(\zeta/s)$, extracted from Figs.~\ref{fig6}(c) and \ref{fig6}(d), plotted as functions of $\langle\mu_{p}\rangle$ using the correspondence between $\sqrt{s_{NN}}$ and $\langle\mu_{p}\rangle$ established in Fig.~\ref{fig7}(a).

The nonmonotonic behavior of these ratios constrains the inflection point to $\langle\mu_{p}\rangle \approx (0.017,0.021)$ GeV, corresponding to $\sqrt{s_{NN}}\approx(39,27)$ GeV.
Thus, the nonmonotonic features in Fig.~\ref{fig7}(b) can be understood in terms of $\langle\mu_{p}\rangle$.
One concludes that thermal-vorticity-induced spin polarization significantly broadens the range of parton chemical potentials sensitive to the QCD equation of state.

\section{Conclusions}
\label{conclusions}
This work employed the AMPT model to simulate the space-time evolution of noncentral O+O collisions.
The results show that including SP significantly affects the EoS, particularly its effective transport and TTCs of partons.
In particular, one finds a nonmonotonic, energy-dependent behavior in the ratios $\tilde{c}_{s}^{2}/c_{s}^{2}$ and $(\tilde{\zeta}/\tilde{s})/(\zeta/s)$, which arises mainly from SP corrections to $\langle dP/d\varepsilon\rangle$.
These results suggest that SP may serve as an effective probe of the QCD EoS.

This study accounts only for the contribution of thermal vorticity to SP, while neglecting other possible sources such as thermal shear and magnetic fields.
Moreover, the analysis is based on kinetic theory and does not model the self-consistent evolution of microscopic dynamics.
A more comprehensive treatment should include these additional effects.

Despite these limitations, the nonmonotonic behavior of the TTCs exhibits an inflection point around $\sqrt{s_{NN}}=27$ GeV, providing constraints on the QCD EoS within the framework of this simplified model.
Thus, these findings may offer useful guidance for future investigations of QGP properties.

\appendix
\section{time- and energy-dependent TTCs}
\label{appendix}

\begin{figure*}[t]
    \centering
    \includegraphics[width=0.80\textwidth]{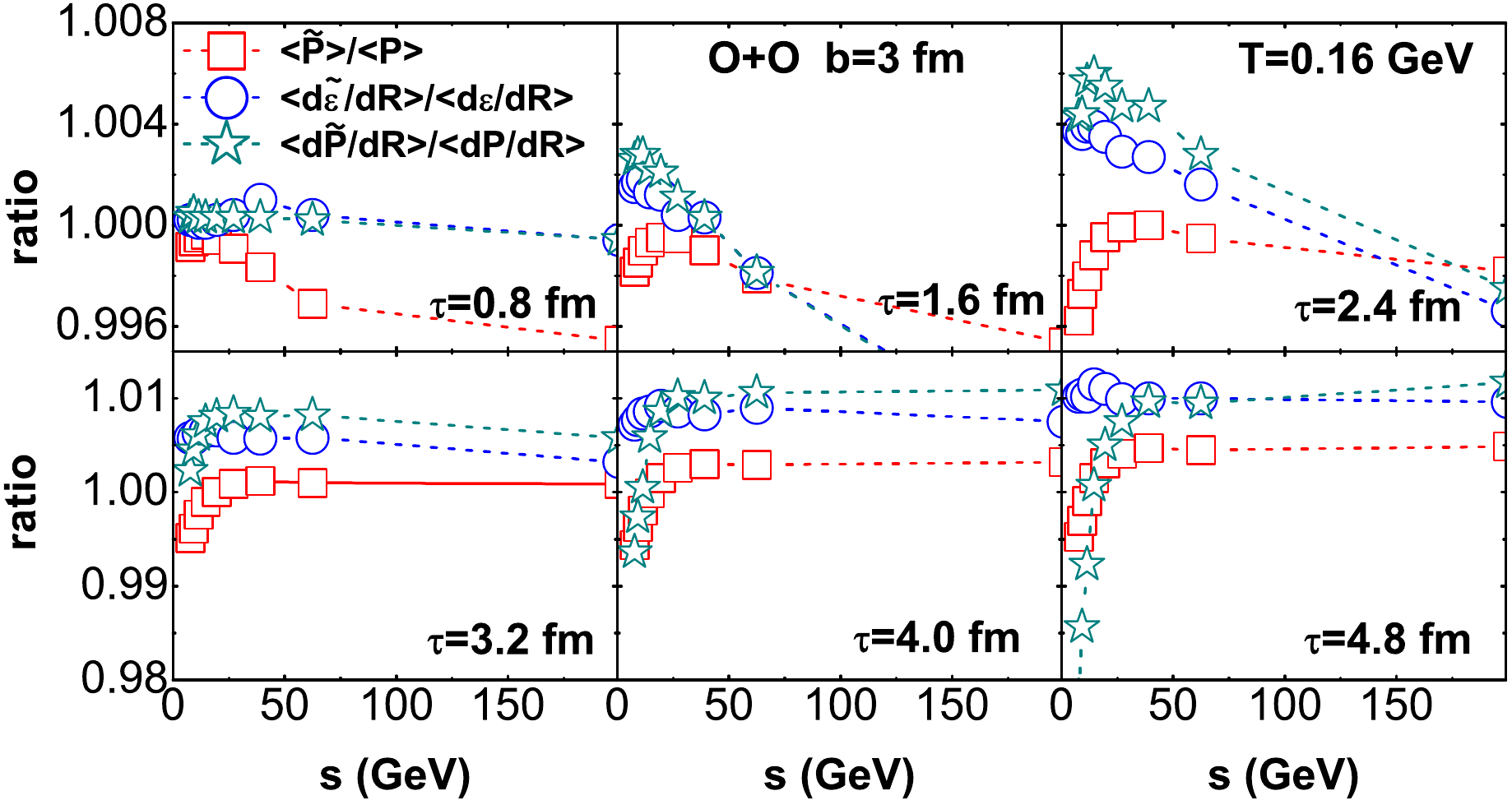}
     \caption{(Color online)
      Comparison of SP and non-SP results.
      The ratios $\langle \tilde{P}\rangle/\langle P\rangle$, $\langle d\tilde{\varepsilon}/dR\rangle/\langle d\varepsilon/dR\rangle$, and
      $\langle d\tilde{P}/dR\rangle/\langle dP/dR\rangle$ with $T=0.16$ GeV, shown as functions of collision energy for O+O collisions in the range $\tau=0.8$--4.8 fm.}
    \label{fig:gradenergy}
\end{figure*}

\begin{figure}[tp]
    \centering
    \includegraphics[width=0.480\textwidth]{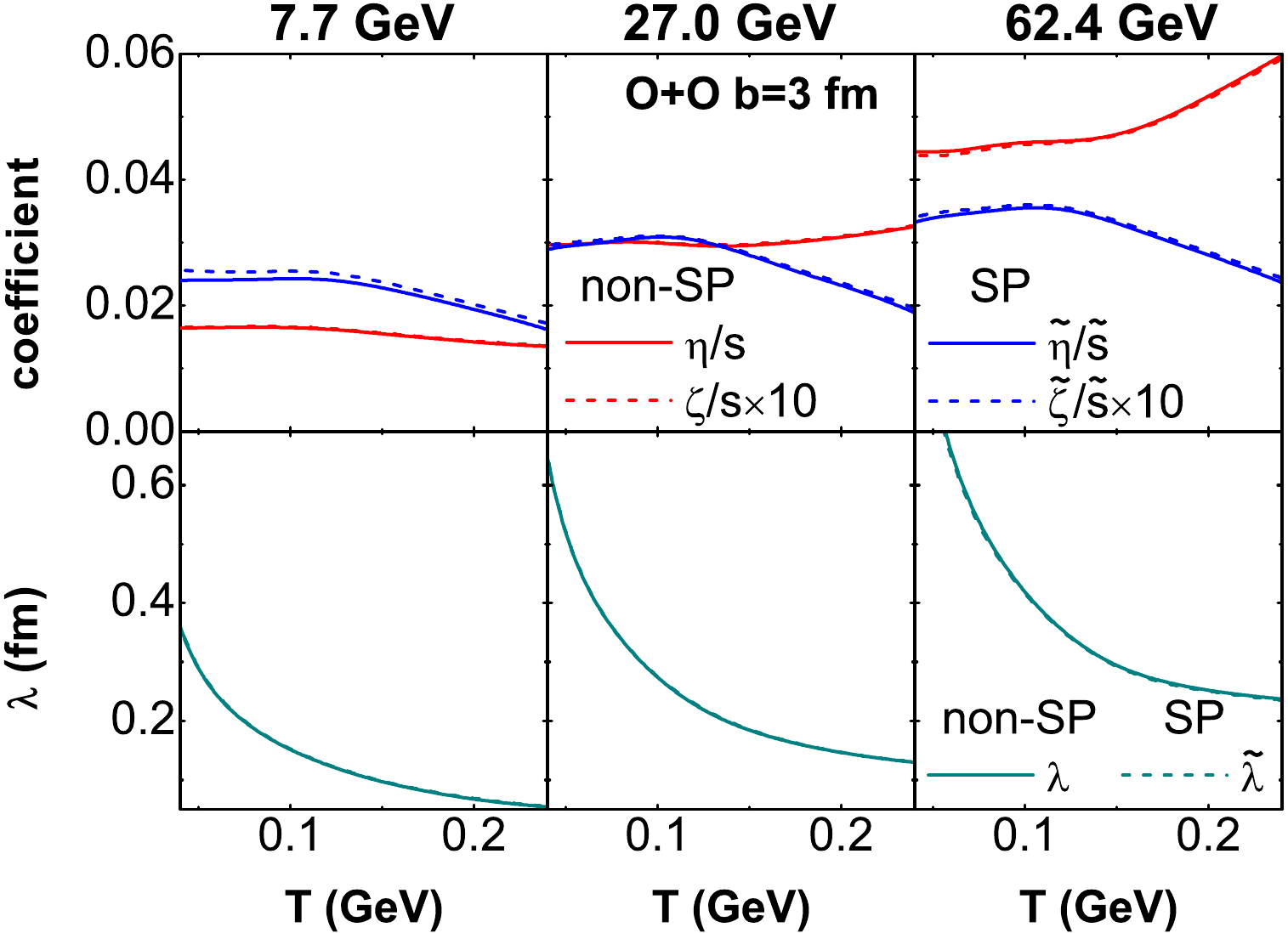}
    \caption{(Color online)
     Comparison of SP and non-SP results.
     The transport and thermodynamic coefficients $\eta/s$, $\zeta/s$, and $\lambda$ are shown as functions of temperature for O+O collisions at $\sqrt{s_{NN}}=7.7$, 27.0, and 62.4 GeV.
     Results are integrated over proper time from $\tau=0.2$ to 5 fm.}
    \label{fig:coeffenergy}
\end{figure}

\begin{figure}[tp]
    \centering
    \includegraphics[width=0.480\textwidth]{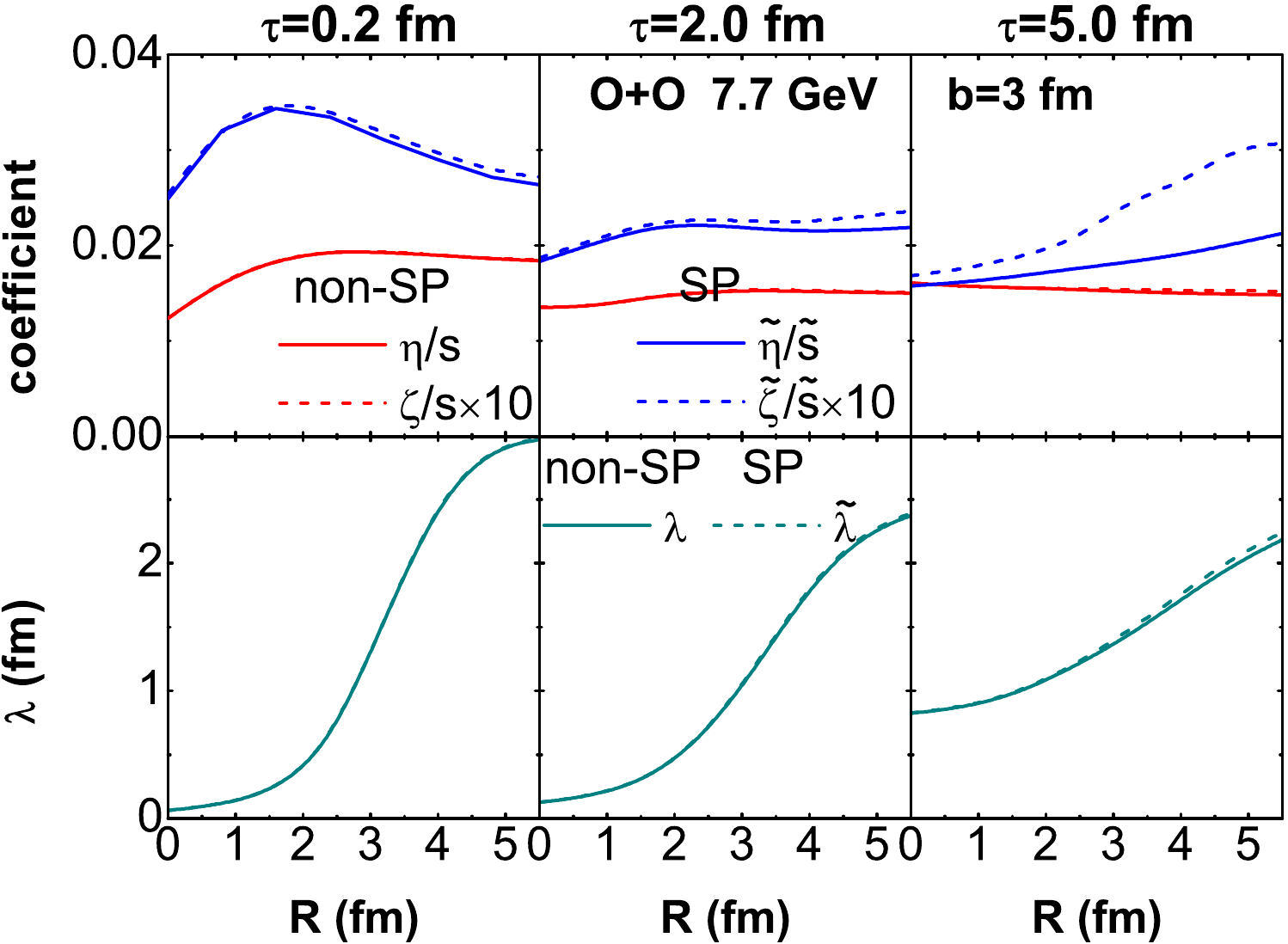}\\
    \includegraphics[width=0.480\textwidth]{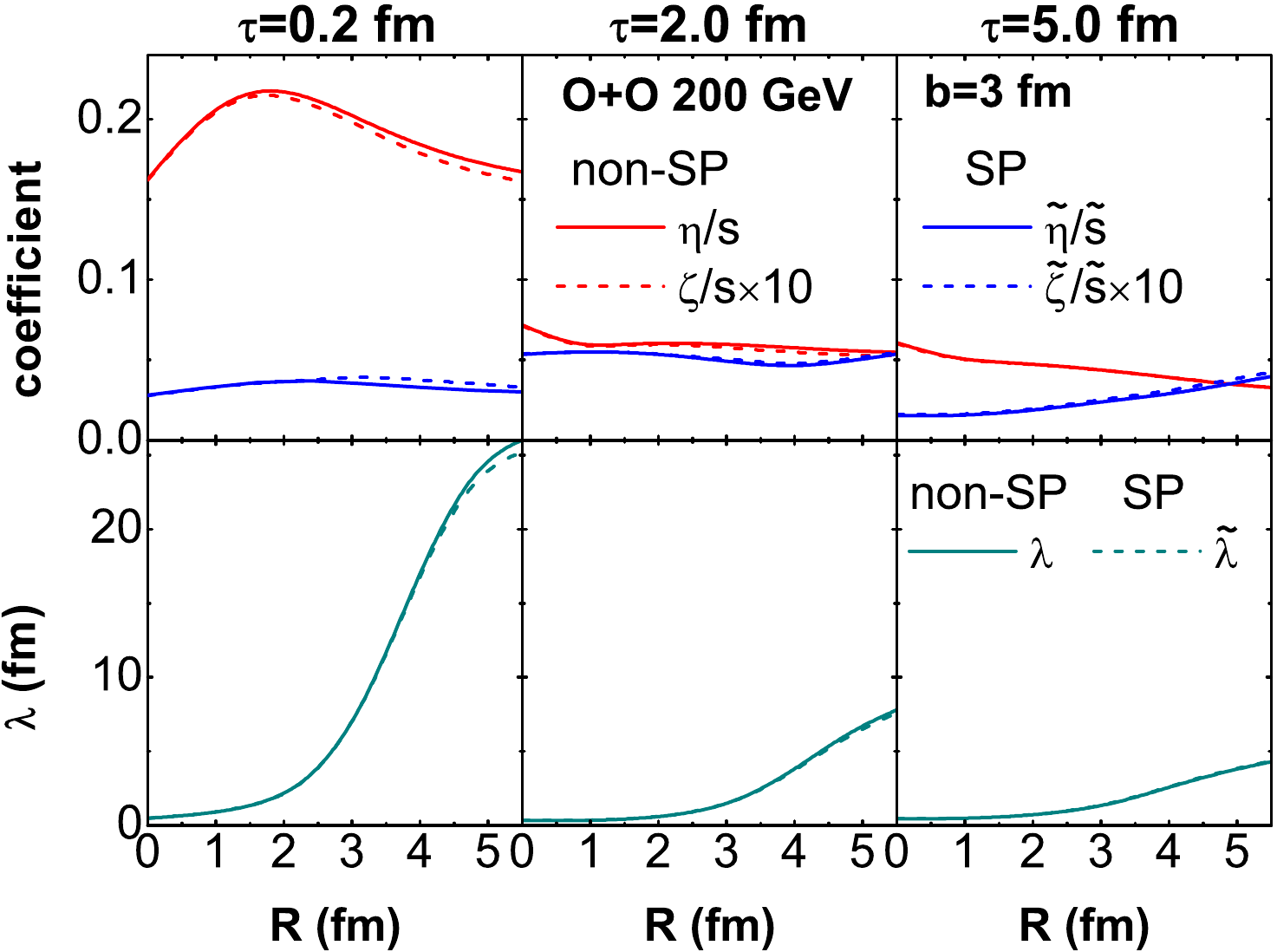}
    \caption{(Color online)
    Comparison of SP and non-SP results.
    The transport and thermodynamic coefficients $\eta/s$, $\zeta/s$, and $\lambda$ are shown as functions of radius for O+O collisions at $\sqrt{s_{NN}}=7.7$ GeV (top panels) and 200 GeV (bottom panels).
    Results are given at proper times $\tau=0.2$, 2.0, and 5.0 fm.}
    \label{fig:coefftime}
\end{figure}

Figure~\ref{fig:gradenergy} shows the ratios $\langle \tilde{P}\rangle/\langle P\rangle$,
$\langle d\tilde{\varepsilon}/dR\rangle/\langle d\varepsilon/dR\rangle$, and
$\langle d\tilde{P}/dR\rangle/\langle dP/dR\rangle$
with $T=0.16$ GeV, for O+O collisions at different proper times.
The differences between SP and non-SP results are most pronounced at early times for higher collision energy, and at later times for lower energy.
As time increases, the effects of SP on the quantities become more pronounced in the low-energy region but remain at unity in the high-energy region.

Figure~\ref{fig:coeffenergy} shows the transport and thermodynamic coefficients---specific shear viscosity $\eta/s$, specific bulk viscosity $\zeta/s$, and mean free path $\lambda$---for O+O collisions at $\sqrt{s_{NN}}=7.7$, 27.0, and 62.4 GeV.
The differences between SP and non-SP results vary significantly across these energies.

Figure~\ref{fig:coefftime} shows the transport and thermodynamic coefficients---specific shear viscosity $\eta/s$, specific bulk viscosity $\zeta/s$, and mean free path $\lambda$---for O+O collisions at $\sqrt{s_{NN}}=7.7$ and 200 GeV at different proper times.
The coefficients exhibit opposite time dependences in the two systems.
In the 7.7-GeV case, the difference between SP and non-SP results is most pronounced at later times for $\zeta/s$ and $\lambda$, while $\eta/s$ show only weak time dependence.
In the 200-GeV case, the differences are largest at early times and gradually diminish as the system evolves.

\begin{acknowledgments}
This work is supported by National Natural Science Foundation of China Grant No.~12105057, and Guangxi
Natural Science Foundation Grants No.~2023GXNSFAA026020 and No. 2019GXNSFBA245080.
\end{acknowledgments}

\bibliographystyle{apsrev4-1}
\bibliography{References}

\end{document}